\documentclass{sig-alternate}
\usepackage{todonotes}

\newtheorem{prob}{Theorem}

\newtheorem{pro}[prob]{Problem}

\begin{document}

\pagenumbering{arabic}

\renewcommand\floatpagefraction{0.99}
\renewcommand\dblfloatpagefraction{0.99} 
\renewcommand\topfraction{0.98}
\renewcommand\dbltopfraction{0.98} 
\renewcommand\bottomfraction{0.98}
\renewcommand\textfraction{.00}   
\setcounter{totalnumber}{50}
\setcounter{topnumber}{50}
\setcounter{bottomnumber}{50}





%

\title{Detecting Hierarchical Ties Using Link-Analysis Ranking at Different Levels of Time Granularity}

\numberofauthors{3} 

\author{
\alignauthor
Hend Kareem\\
       \affaddr{Stockholm University}\\
       \affaddr{Stockholm, Sweden}\\
       \email{hindoo86@googlemail.com}
\alignauthor
Lars Asker\\
       \affaddr{Stockholm University}\\
       \affaddr{Stockholm, Sweden}\\
       \email{asker@dsv.su.se}
\alignauthor Panagiotis Papapetrou\\
       \affaddr{Stockholm University}\\
       \affaddr{Stockholm, Sweden}\\
       \email{panagiotis@dsv.su.se}
}
\additionalauthors{Additional authors: John Smith (The Th{\o}rv{\"a}ld Group,
email: {\texttt{jsmith@affiliation.org}}) and Julius P.~Kumquat
(The Kumquat Consortium, email: {\texttt{jpkumquat@consortium.net}}).}
\date{30 July 1999}

\maketitle

\begin{abstract}
Social networks contain implicit knowledge that can be used to infer hierarchical relations that are not explicitly present in the available data. Interaction patterns are typically affected by users' social relations. 
We present an approach to inferring such information that applies a link-analysis ranking algorithm at different levels of time granularity. In addition, a voting scheme is employed for obtaining the hierarchical relations.  The approach is evaluated on two datasets: the Enron email data set, where the goal is to infer manager-subordinate relationships, and the Co-author data set, where the goal is to infer PhD advisor-advisee relations.  The experimental results indicate that the proposed approach outperforms more traditional approaches to inferring hierarchical relations from social networks.

\end{abstract}



%

\section{Introduction}
\label{sec:first}
Social ties are useful for understanding the structure of a social network. They can be used for link prediction \cite{kumariea2014}, rating prediction, product recommendation \cite{sun2011co}, and community discovery \cite{parthasarathy2011community}. Some types of social ties, like friendship relations in Facebook, are explicit, while others are implicit. Furthermore, social ties might be hierarchical, such as manager-subordinate ties between employees in a company or advisor-advisee between co-authors in a scientific publication community. According to Jaber et al. \cite{Jaber:2014:IOH:2567948.2580070}, inferring such ties plays a vital role in classifying actors in a network and discovering different communities. Sometimes influential actors in a network can also be detected using hierarchical social relationships. These hierarchical ties can be further used to validate social and psychological theories, such as "Opinion Leader" using a two-steps theory suggesting that ideas first flow to "opinion leaders" and then to "ordinary users" \cite{Jaber:2014:IOH:2567948.2580070}\cite{Tang:2012:IST:2124295.2124382}. Furthermore, inferring advisor-advisee, for example, can help in studying the evolution of research history in different research communities' relations and understanding the influence of a researcher in these communities \cite{Wang:2010:MAR:1835804.1835833}.

Much research on social networks has focused on inferring attributes of users based on the characteristics of other users in the same network \cite{mislove2010you}; or the opposite, i.e., extracting the structure of a social network by using users' characteristics \cite{Adamic2003211}. Another line of research has focused on labeling the ties between actors in a network and measuring the strength or defining the directions of undirected ties \cite{Yang:2012:FFP:2348283.2348359} \cite{mcpherson2001birds} \cite{Adali:2012:ASL:2187836.2187930} \cite{Xiang:2010:MRS:1772690.1772790} \cite{Zhang:2014:PRR:2566486.2567968} \cite{sintos2014using}.  Nonetheless, limited attention has been given to the role that ``time" plays in inferring implicit hierarchical ties among a set of users in a network. 

\smallskip
\noindent
\textbf{Contributions.} The main contribution of this paper is to
address the problem of inferring hierarchical ties in a social network by employing link-analysis ranking at different levels of time granularity. The proposed solution is compared against two real-world datasets showing competitive performance against a baseline competitor method.   Our solution extends the work by Jaber et al. \cite{Jaber:2014:IOH:2567948.2580070} as follows: (1) we employ a time-based technique that explores the social graph at different levels of time granularity; (2)
we explore both weighted and unweighted representations of the social graphs, and demonstrate the benefits of the latter.

\section{Related Work}
An extensive amount of research has been performed on inferring the type and the strength of social ties in a network. Sintos and Tsaparas \cite{sintos2014using} examine the problem of labeling connections in a network depending on whether they are strong or weak using only the graph structure of the network. Similarly, Xiang and colleagues  \cite{Xiang:2010:MRS:1772690.1772790} estimate relationships' strength by considering interaction activity and user similarity based on the Homophily theory in order to develop an unsupervised model that exhibits the latent properties of the network.
Adali et al. \cite{Adali:2012:ASL:2187836.2187930} utilize the statistical properties of communication patterns among actors to deduce the type and strength of links in a network. Backstrom and Kleinberg \cite{Backstrom:2014:RPD:2531602.2531642} use the underlying network structure to identify the most influential person in an actor's social network neighborhood.
Yang et al. \cite{Yang:2012:FFP:2348283.2348359} focus on the problem of labeling the edges in a social network as positive 
or negative 
based on the user behavior of decision-making. 

More recently, Zhang et al. \cite{Zhang:2014:PRR:2566486.2567968} argue that not only the sign 
and strength 
are important when it comes to obtaining a better understanding of social network structure, but also the direction of the ties between the actors. Liebowitz \cite{tsui2005linking} integrates the usage of an analytical hierarchical process with social network analysis on the organizational level to create a knowledge map.  Moreover, Gupte et al. \cite{Gupte:2011:FHD:1963405.1963484} adopt the principle of social stratification in their approach, which refers to the categorization of people in society into ranked groups based on their status, power, wealth or knowledge. By applying stratification on humans, they assumed that people who are higher up in the hierarchy tend to have a higher status (ranking) in comparison to people who are lower in the hierarchy. Further, people at the top levels of the hierarchy are less likely to connect to people at low levels of the hierarchy. An algorithm has been suggested to find the best hierarchy in a directed network. However, they have not considered the time dimension in the problem. Moreover, they study the problem at a network-level and not at an actor-level as in this study.

Finally, link prediction plays an essential role in discovering interactions within social networks. Along this line, it draws an immense interest in the field of data mining and networks communication. The link prediction problem is one of the underlying problems in social network evaluation. It has been used in different contexts, for instance, in companies to discover the interactions within social networks \cite{Liben-Nowell:2003:LPP:956863.956972}, even in security sector by monitoring terrorist networks \cite{huang2009time} or to be used in prediction of missing links in a community. 

\section{Detecting Hierarchical Ties}
In this study, we approach the problem of inferring hierarchical ties in a social network as a ranking problem.  

\smallskip
\noindent
\textbf{Problem Formulation.} More formally, a social network can be represented as a graph $G = (V, E^c, E^s, W)$, with:
\begin{itemize}
\item $V = \{v_1,v_2,... v_n\}$: the set of nodes representing the actors in a network;
\item $E^c$: the set of edges corresponding to the interactions between the actors in a network;
\item	$E^s \subseteq V \times V$: the set of hierarchical relationships between the actors in $G$. Each element $(u, v) \in E^s$ is a pair of nodes, where $u$ is the direct superior of $v$;
\item	$W$: is a vector of edge weights, with $w_uv \in W$ being the weight of the edge connecting nodes $u,v$. 
\end{itemize}
The problem this study addresses is as follows:
\begin{pro}
Given a social graph $G (V, E^c, W)$ and a set of nodes $Q$, our aim is to infer the set of direct hierarchical ties of each node in $Q$ using $G_{output} (V, E^s)$.
\end{pro}

For instance, given a set of e-mails exchanges between employees in a company, or co-authored scientific papers, the goal is to infer direct hierarchical ties such as, manager-subordinate, or advisor-advisee, respectively.

\smallskip
\noindent
\textbf{Solution.} We approach the problem based on the ``Opinion Leader" theory, which is based on the assumption that in a social network ideas flow from opinion leaders to ordinary users. Hence, opinion leaders are the most influential members and have significant impact on the other users \cite{Tang:2012:IST:2124295.2124382}. One way of capturing this impact is by using link-analysis ranking \cite{Jaber:2014:IOH:2567948.2580070}.  Accordingly, opinion leaders will tend to have a higher ranking score, such as managers or advisors, than ordinary users, such as subordinates and advisees. 

\smallskip
\noindent
\textbf{Baseline Approach.} One solution is to employ Rooted-PageRank as a standard link-analysis ranking method. The main idea behind Rooted-PageRank is to calculate the importance scores of each node in the graph relative to a predefined root node. Rooted-PageRank can be directly adopted for our problem by setting the root to be the actor whose direct hierarchical tie is to be identified. The node with the strongest connection to the root receives the highest Rooted-PageRank score, and hence corresponds to the actor with the strongest direct hierarchical tie. This constitutes the baseline approach (proposed by Jaber et al. \cite{Jaber:2014:IOH:2567948.2580070}).

\smallskip
\noindent
\textbf{Time-based Approach.}  In this study, we investigate a time-based solution using Rooted-PageRank. More specifically, the total time span [$t_{start}$,$t_{end}$] of the interactions in the graph is divided into $m$ equal-sized non-overlapping time slots. For each time slot $t^k=[t^k_i,t^k_j]$, we define subgraph $G_k$  = ($V_k$, $E_k$, $W_k$), where $V_k$ is the set of actors who interact with at least one other actor within time slot $t^k$, $E_k$ is the set of edges between the actors at time slot $t^k$, and $W_k$ is the vector of edge weights corresponding to that time slot. In this study we explore two versions of Rooted-PageRank: unweighted, where edge weights are binary  indicating whether two nodes have at least one interaction during $t^k$ (indicated with a weight of $1$), and weighted, where each edge weight corresponds to the total number of interactions between the adjacent nodes during $t^k$.

Our method proceeds by applying Rooted-PageRank to each subgraph $G_k$.  The algorithm follows three steps:
\begin{enumerate}

\item \textbf{Rank: } For each node $v\in G_t$, we run Rooted-PageRank having set $v$ as the root. This produces a set of scores $\mathcal{R}(v) = \{RS(u_i)\}$, $u_i \in V_k\setminus v$ and $RS(u_i)$ denoting the Rooted-PageRank score of node $u_i$.

\item \textbf{Sort: } Each $\mathcal{R}(v)$ is sorted in descending order, resulting in a sorted list $L^k(v)$ for each $v\in G_k$.

\item \textbf{Merge: } For each node $v\in G$, i.e., each node in the original graph, the corresponding sorted lists are aggregated as follows: given an integer threshold $p$, the aggregate score of node $v$ is the number of times it appears in a position within $1,p]$ in $L^k(v),\forall k\in [1,m]$.
\end{enumerate}
The final ranking is inferred by the resulting aggregation.

\section{Experimental Evaluation}
\subsection{Setup}

Experiments have been performed on two real datasets:
\begin{itemize}
\item
{\sc Enron}: includes more than 255000 emails sent between 87474 users. A subset was used consisting of all emails sent between any of the 155 email addresses of Enron employees, 146 of them having a known direct superior. The type of interaction is email, and the type of hierarchical tie to be identified is manager-subordinate. Each record in the data set describes a directed link in the network consisting of sender, receiver, and the total number of emails from sender to receiver per week between January 2000 and November 2001. The dataset was modeled both as an unweighted graph (edges were assigned with either 0 or 1 depending on whether there is at least one email sent from one node to the other) and as a weighted graph (edge weights are proportional to the number of emails that are sent between the corresponding pair of adjacent nodes). Two different time slots were investigated for this dataset, week and month.
\item
{\sc Co-author}: includes more than 1 million authors involved in 80000 papers between 1967 and 2011. The type of hierarchical tie of interest is the advisor-advisee, which is known for 2098 authors in this dataset.  In contrast with {\sc Enron}, the co-author relationship is symmetric, hence the graph is undirected. We have selected a subset consisting of 700 authors who are advisees, for which the advisor is known, while the first and last year of publication is in the range between 2001 and 2010, together with all co-authors, resulting in 2136 authors altogether. 
\end{itemize}
Both versions of our method, unweighted and weighted, are benchmarked against the competitor baseline in terms of recall, which is computed as follows: for each rank $i$ we compute the percentage of nodes whose direct hierarchical tie appears in a rank equal to or higher than rank $i$.

\subsection{Results}

\smallskip
\noindent
\textbf{Results on Enron.}  For the {\sc Enron} dataset both the baseline approach and the time-based approach are assessed. For each approach, both weighted and unweighted graphs are studied. Finally, for the time-based approach both the time slot per week and per month are examined.

\begin{figure}[ht]
\includegraphics[width=0.9\columnwidth]{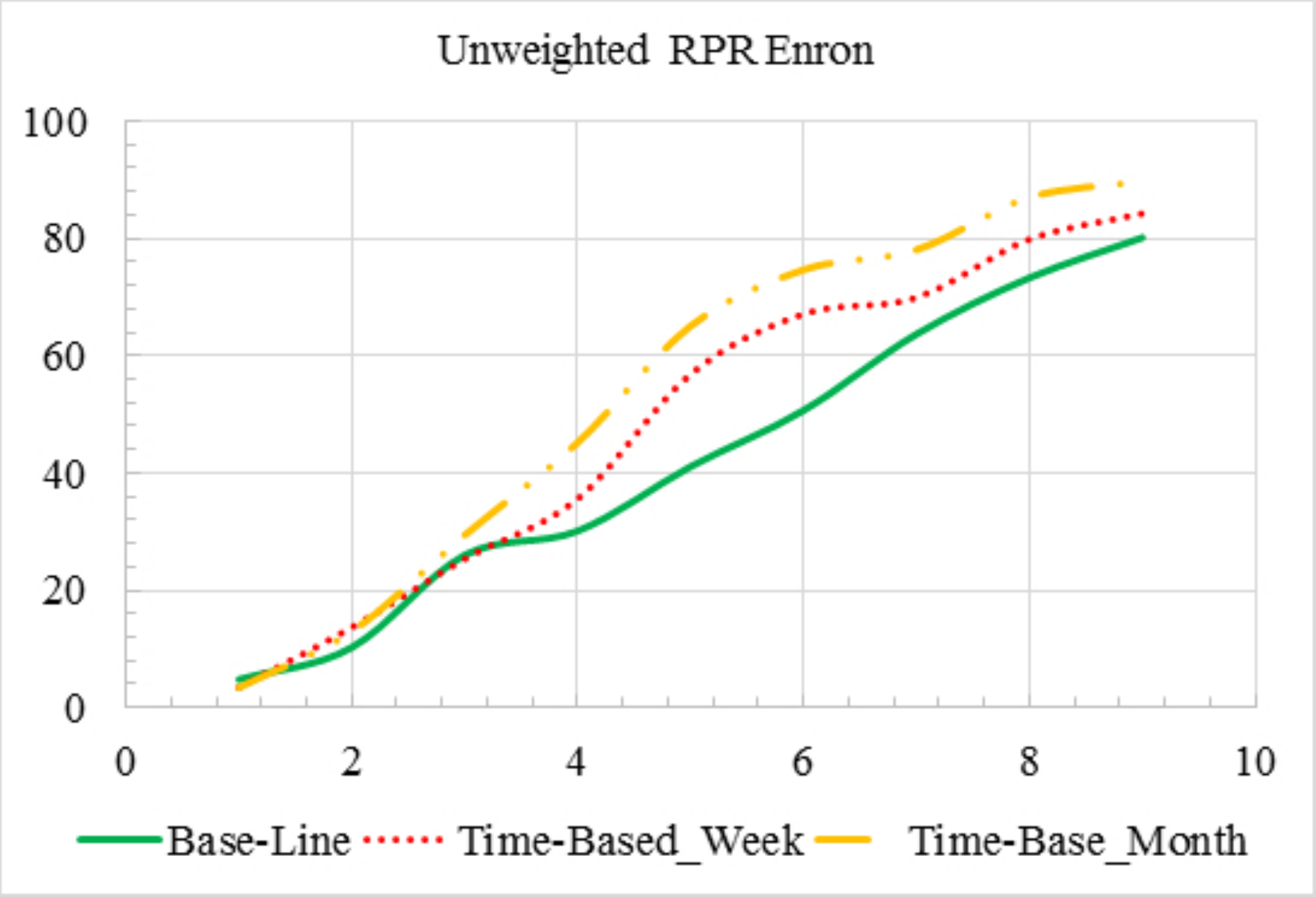}
\caption{The comparison of the results for the unweighted RPR approaches for the Enron dataset. Y-axis shows the percentage of advisors that appear within the rank given by the X-axis.}
\label{fig:fig6}
\end{figure}

\begin{figure}[ht]
\includegraphics[width=0.9\columnwidth]{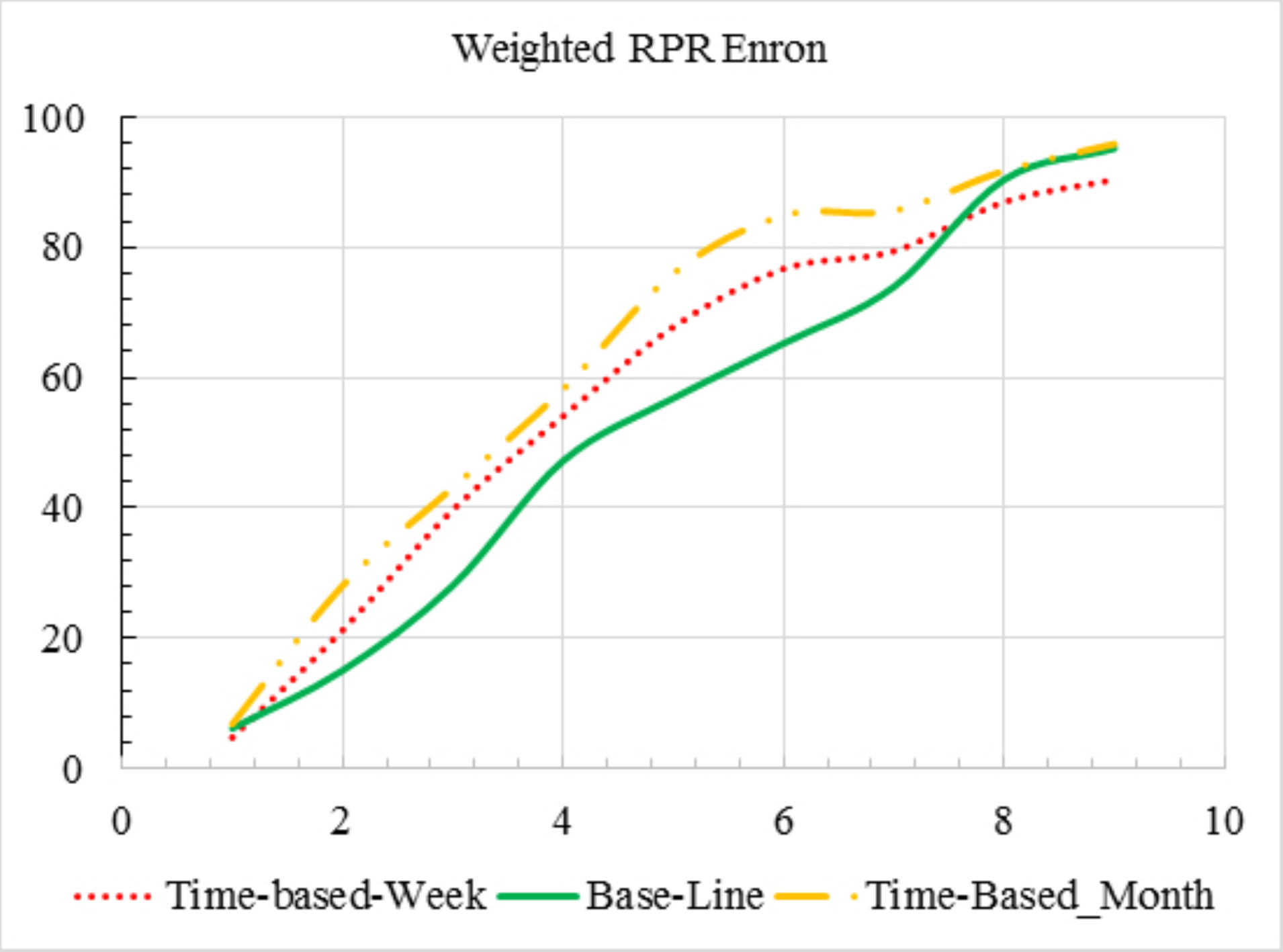}
\caption{The comparison of the results for the weighted RPR approaches for the Enron dataset. Y-axis shows the percentage of advisors that appear within the rank given by the X-axis.}
\label{fig:fig7}
\end{figure}

These results show that the weighted RPR approaches outperform the unweighted RPR for all approaches with and without time consideration and with different time precisions.  Figures \ref{fig:fig6} and \ref{fig:fig7} show that the time precision per month performed better comparing to the time precision when the time slot is per week for both the weighted and the unweighted RPR algorithms. Further, the approaches that take the time into consideration perform better than the baseline method where no time dimension is considered.

\smallskip
\noindent
\textbf{Results on Co-author.}  Both the baseline and the time-based approaches are evaluated. For each one of them both the unweighted and the weighted undirected graphs are built. When it comes to the time precision, for this network only one time precision is considered, namely per year. As stated previously, for the baseline approach no temporal aspect is considered for the interactions. 

\smallskip
\noindent
For the Weighted RPR approach, the number of papers co-authored is taken into consideration, and the weights assigned according what has been described above. These results indicate that the weighted RPR performs better than the unweighted PRP for the baseline approach where no time is considered.

For the Time-Based approach, a separate graph is built for each year between 2001 and 2010.  Again the position of the advisors that occurs most frequently is taken as the position suggested for the supervisor for the given advisee. For this approach both the unweighted and the weighted RPR are considered in order to be able to compare the performance in terms of the recall.

\begin{figure}[ht]
\includegraphics[width=0.9\columnwidth]{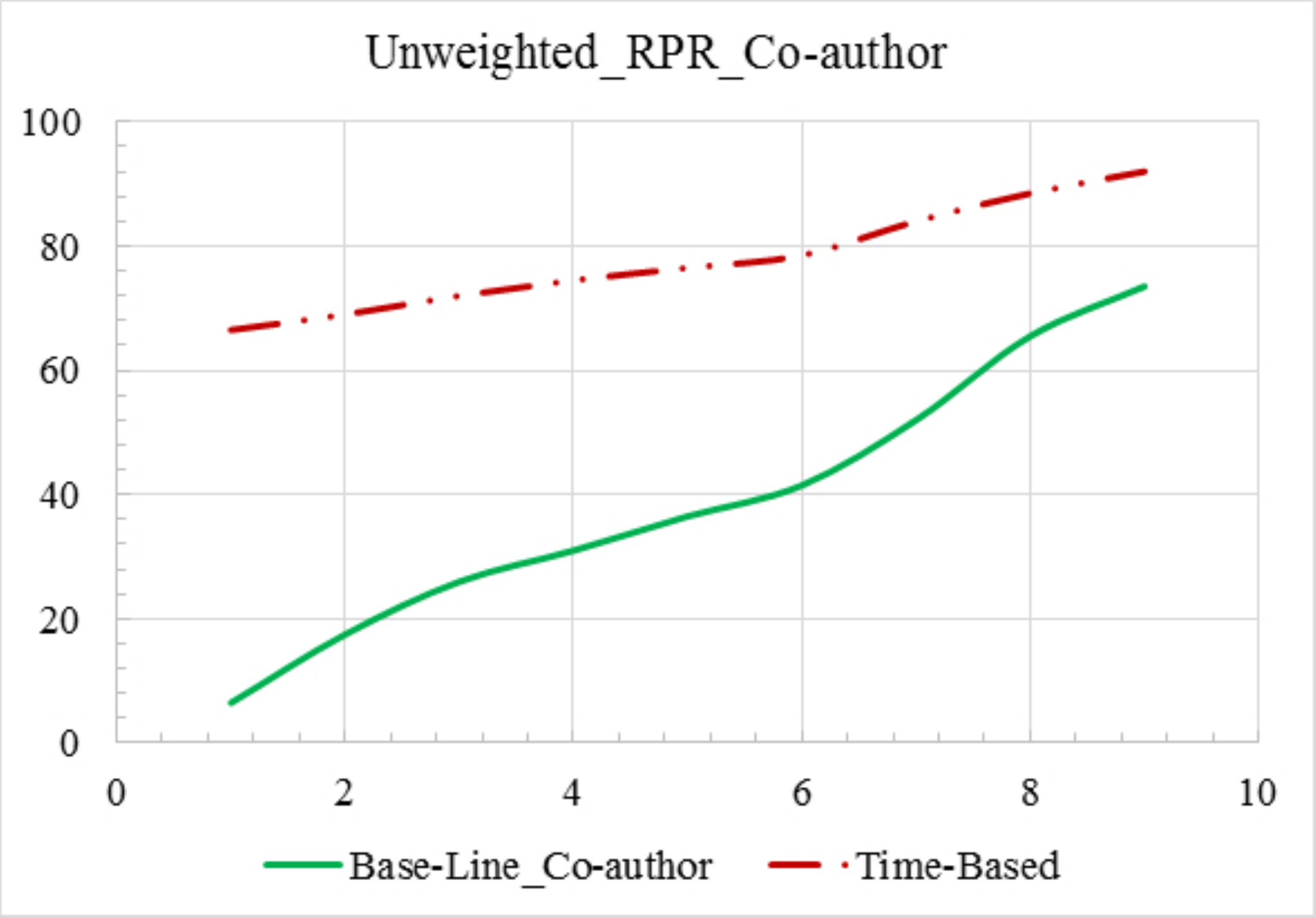}
\caption{Comparison of the results for the unweighted RPR approaches for the Co-author dataset. Y-axis shows the percentage of supervisors that appear within the rank given by the X-axis.}
\label{fig:fig10}
\end{figure}

\begin{figure}[ht]
\includegraphics[width=0.9\columnwidth]{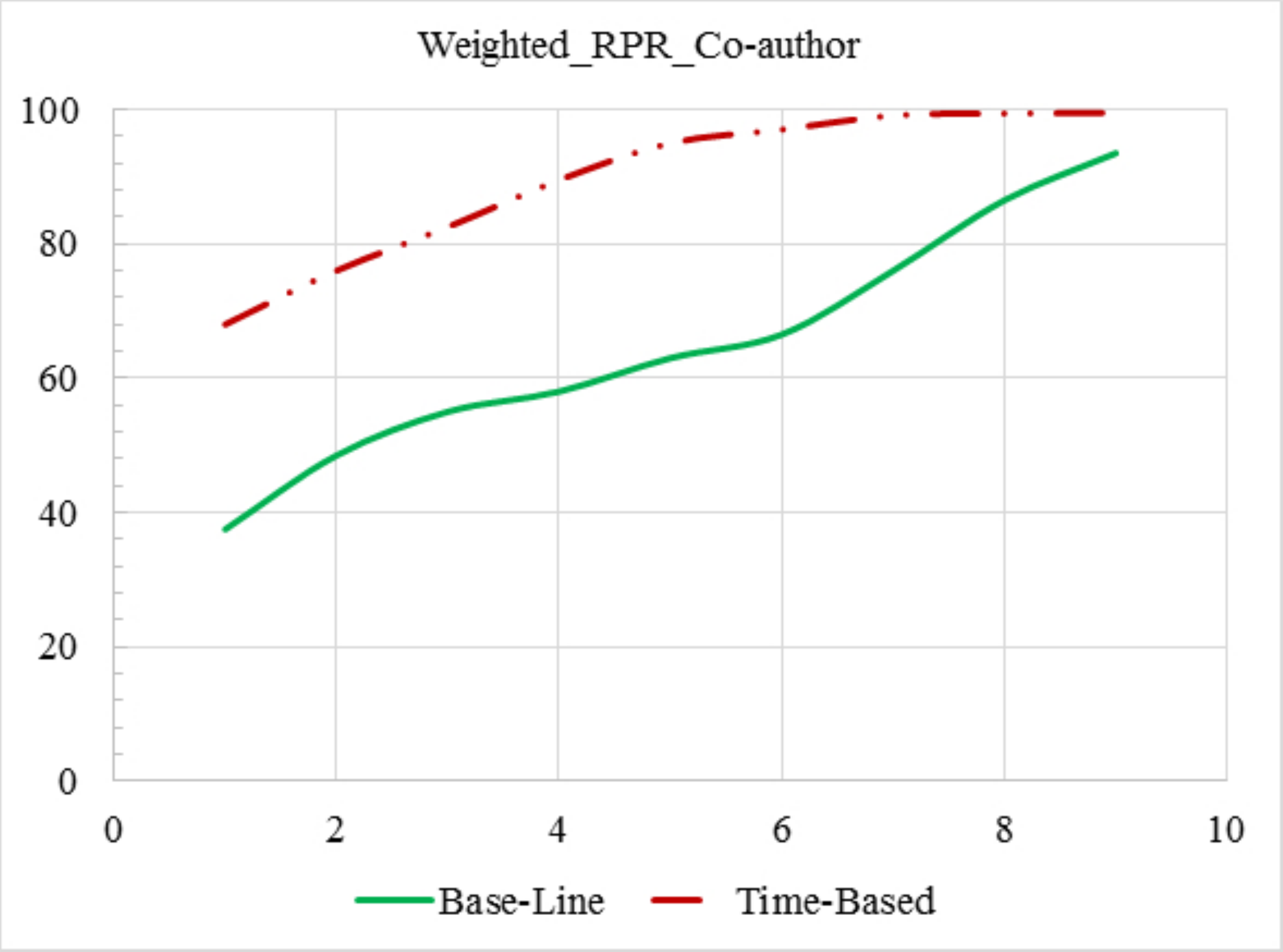}
\caption{Comparison of the results for the weighted RPR approaches for the Co-author dataset. Y-axis shows the percentage of supervisors that appear within the rank given by the X-axis.}
\label{fig:fig11}
\end{figure}

\smallskip
\noindent
\textbf{Overall Comparison.}  When comparing the results for the weighted and unweighted approaches for both the baseline and the time-based it can be noted that the weighted approaches outperform the unweighted methods. The resulted scatter plots are shown in Figures \ref{fig:fig10}, \ref{fig:fig11}, \ref{fig:fig7}, \ref{fig:fig6}. For the Co-author dataset, the time plays a crucial role in the task of inferring the advisor-advisee relationship for both the unweighted and the weighted RPR.

\section{Conclusions}

We  proposed  a time-based link analysis ranking based approach for inferring  direct hierarchical ties in social graphs. The key novelty of our approach is to exploit the link information in the graph at different time granularity levels, and employ a final voting scheme for obtaining the ties. Our findings indicate that the proposed approach outperforms the baseline competitor method in terms of recall on two real datasets.  In addition, the consideration of weighted edges instead of unweighted (binary) representations yields higher recall in both datasets. Future work includes the consideration of potential temporal dependencies between the involved interactions and the investigation of the underlying event interaction distributions.
\bibliographystyle{plain}
\bibliography{paper}

\end{document}